\documentclass[useAMS,usenatbib]{mn2e}
\usepackage{graphicx}
\usepackage{epsfig}

\begin{document}

\title{Calibrating Emission Lines as Quasar Bolometers}

\author[Brian Punsly and Shaohua Zhang]{Brian Punsly and Shaohua Zhang\\
4014 Emerald Street No.116, Torrance CA, USA 90503 and \\
ICRANet, Piazza della Repubblica 10 Pescara 65100, Italy,\\
E-mail: brian.punsly@verizon.net or brian.punsly@comdev-usa.com\\
\\
CAS Key Laboratory for Research in Galaxies and Cosmology,\\
Department of Astronomy, University of Sciences and Technology of China,\\
Hefei, Anhui, 230026, China}

\maketitle \label{firstpage}
\begin{abstract}
Historically, emission lines have been considered a valuable tool
for estimating the bolometric thermal luminosity of the accretion
flow in AGN, $L_{bol}$. We study the reliability of this method by
comparing line strengths to the optical/UV continuum luminosity of
SDSS DR7 radio quiet quasars with $0.4<z<0.8$. We find formulae for
$L_{bol}$ as a function of single line strengths for the broad
components of H$\beta$ and Mg II, as well as the narrow lines of [O
III] and [O II]. We determine the standard errors of the formulae
that are fitted to the data. Our new estimators are shown to be more
accurate than archival line strength estimations in the literature.
It is demonstrated that the broad lines are superior estimators of
the continuum luminosity (and $L_{bol}$) with $H\beta$ being the
most reliable. The fidelity of the each of the estimators is
determined in the context of the SDSS DR7 radio loud quasars as an
illustrative application of our results. In general, individual
researchers can use our results as a tool to help decide if a
particular line strength provides an adequate estimate of $L_{bol}$
for their purposes. Finally, it is shown that considering all four
line strength, simultaneously, can yield information on both
$L_{bol}$ and the radio jet power.
\end{abstract}

The primary measure of the strength of an active galactic nucleus (AGN) is its bolometric luminosity, $L_{bol}$,
the broadband thermal luminosity from IR to X-ray. The
characteristic signature of the thermal component is the "big blue
bump" a large blue/UV excess in the spectral continuum \citep{sun89}. Typically, one does
not have complete frequency coverage of the broadband continuum and
$L_{bol}$ can only be estimated. The most common estimators involve
single point rest frame optical or UV luminosity as an approximation
to the big blue bump, \citet{kas00}, or the luminosity of the optical
continuum as in \citet{mil92}. However, there are many
instances in which the optical/UV continuum is not directly
observable or is contaminated with other sources of flux. Thus, astrophysicists need a surrogate for the continuum
luminosity. The two most prevalent situations that are encountered
are firstly, blazars in which the synchrotron optical/UV emission from
the relativistic jet is Doppler boosted to a brightness that either swamps
the quasar thermal emission or contributes an unknown fraction of the total observed
optical/UV flux. The other common occurrence is when dusty molecular gas
(e.g., the "dusty torus") obscures the thermal
emission produced by the quasar accretion flow from our view. The latter
case is believed to be representative of NLRGs
(narrow line radio galaxies) and Seyfert 2 galaxies \citep{ant93}. Historically, broad line
strengths have been used as a surrogate for continuum luminosity in
blazars \citep{cao01,cel93,cel97,ghi11,guu09,mar03,wan04}. For NLRGs,
the broad line region is not visible by definition and narrow
emission line strengths have been used as a surrogate for the
continuum luminosity \citep{raw91,wil99}.
\par These relationships
between line strength and $L_{bol}$ have never been carefully calibrated or
scrutinized for their accuracy. This Letter intends to do both with
a carefully selected, large sample of SDSS DR7 radio quiet quasars ($0.4<z<0.8$) in which two
prominent narrow lines ([OIII] $\lambda$5007 and [OII]
$\lambda$3727) and two broad lines H$\beta$ $\lambda$4861 and Mg II
$\lambda$2798 are observable for all objects. This allows us to compare
the various derived estimators in the context of a single sample of objects,
which removes the uncertainties associated with sample selection biases if a different set of objects
is used for the calibrations of each individual line strength based estimator. In section 4, we consider these estimators in the context
of the radio loud subsample of the SDSS DR7 quasars ($0.4<z<0.8$).
\begin{figure*}
\includegraphics[scale=0.33]{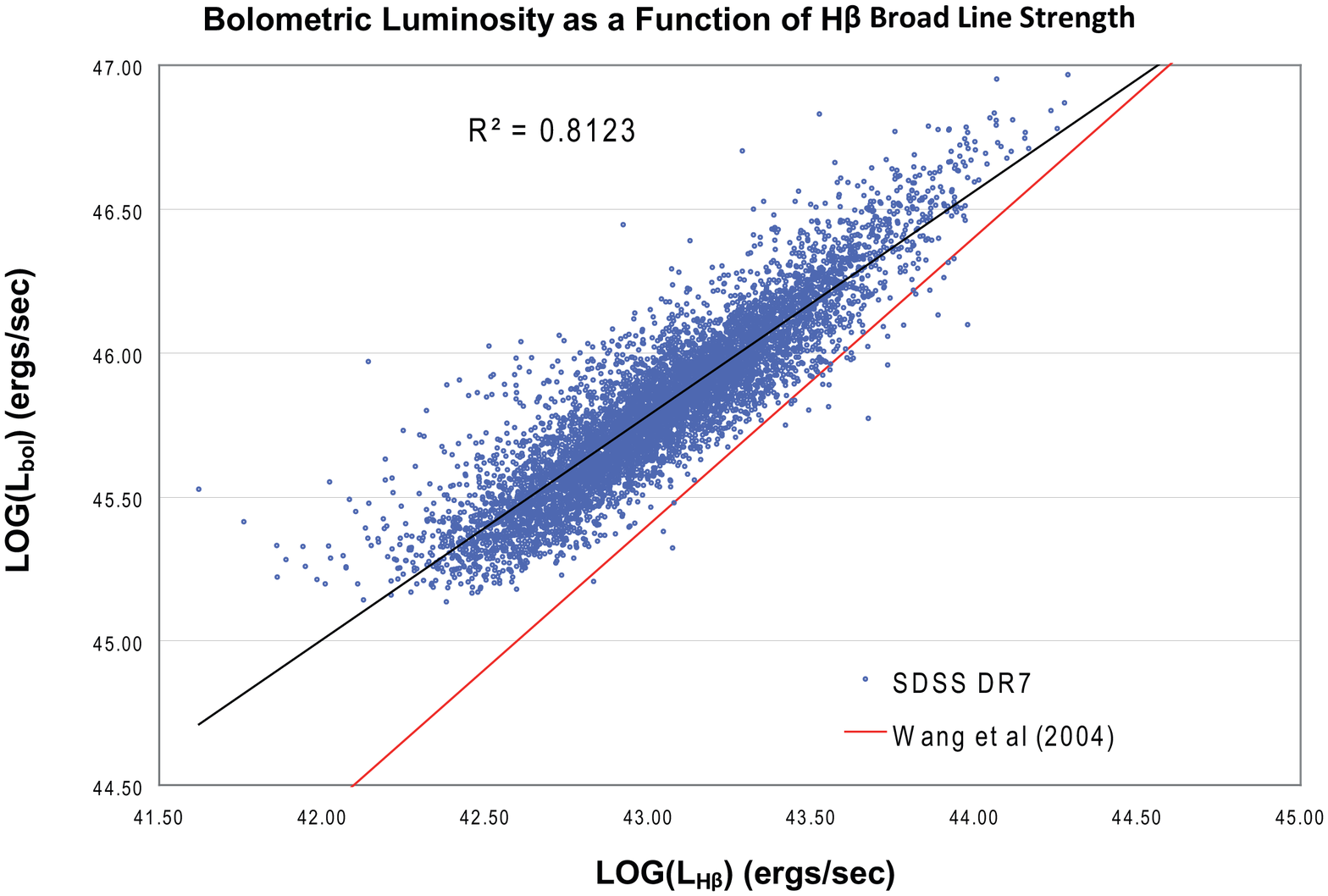}
\includegraphics[scale=0.33]{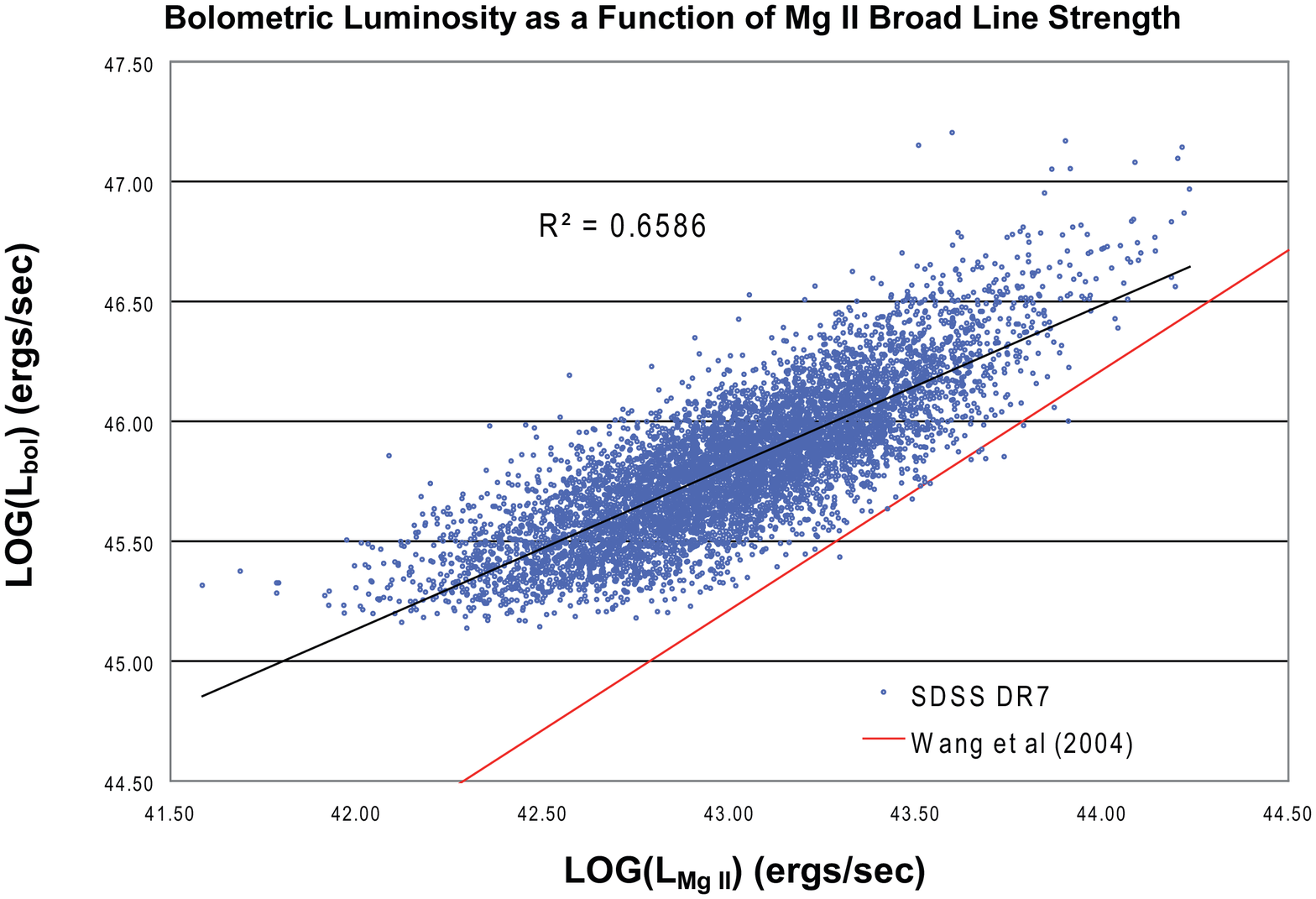}\\
\includegraphics[scale=0.33]{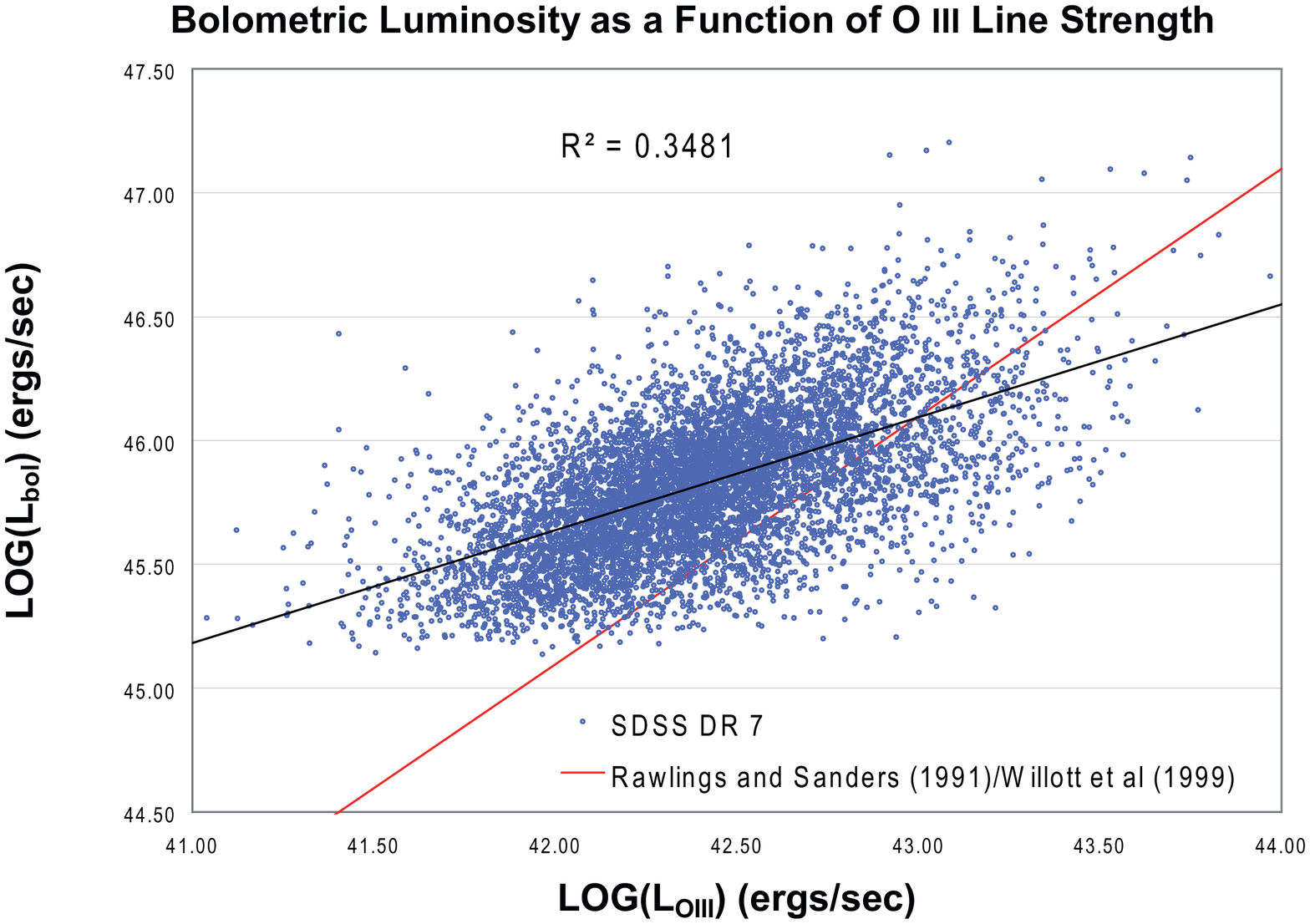}
\includegraphics[scale=0.33]{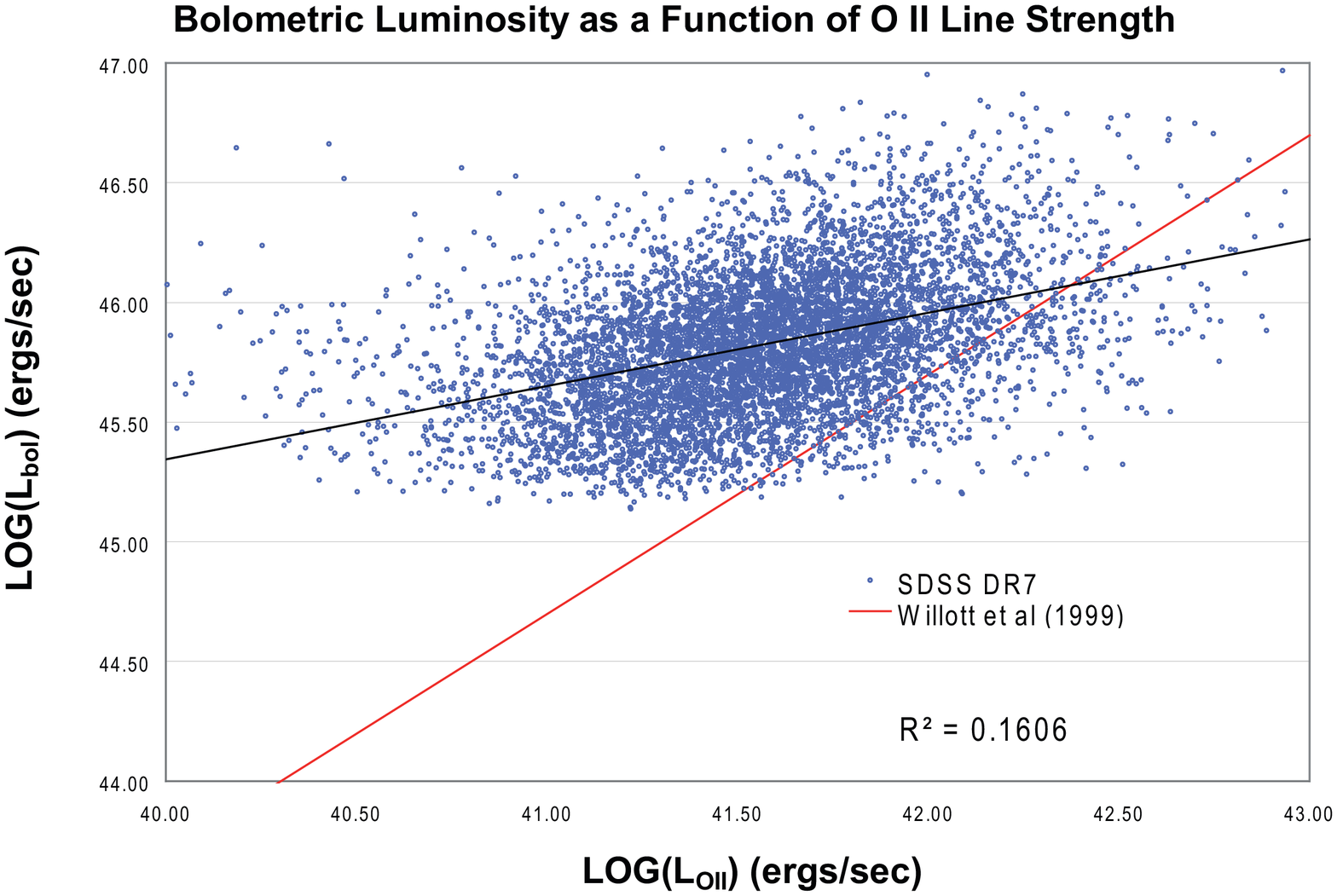}
\caption{Scatter plots of the logarithm of the line strengths as a function of the logarithm of $L_{\mathrm{bol}}=15
L_{\mathrm{cont}}$, where $L_{\mathrm{cont}}$ is the optical/UV
continuum luminosity from 5100 $\AA$ to 3000 $\AA$. The fit to the
SDSS DR7 data is given by the black line. The accuracy of the linear
representation of the data is given by the coefficient of
determination, $\mathbf{R}^{2}$, that is displayed on each plot. An
archival estimator from the literature is shown as a red line.
Notice that the broad component of H$\beta$ is an excellent
representation of $L_{\mathrm{cont}}$ at the top left. The next best
estimator is based on the broad component of Mg II in upper right
hand frame. The [OIII] $\lambda$5007 based estimator is clearly
superior to [OII] $\lambda$3727 based estimator in the bottom row.}
\end{figure*}
\section{Sample Selection}
In order to determine the dependence of line luminosity on the
thermal continuum in a quasar, we constructed a sample of SDSS DR7
radio quiet quasars with $0.4<z<0.8$ and spectra with a
median S/N $>$ 7, this yielded 10069 AGNs. Long slit spectroscopy of radio loud AGN
often show strong regions of narrow line emission on scales as large as $\sim$ 100
kpc that tend to be aligned with the jet direction \citep{ink02,bes00,mcc95}.
The magnitude of this contribution to the narrow line luminosity is largely
unknown \citep{wil99}. Therefore, we segregated out the radio loud quasars because there is a
concern that jet propagation can enhance the line strengths.
The SDSS DR7 data was cross-referenced to
the FIRST data base. The radio loudness, $R$, is usually defined as
a 5 GHz flux density 10 times larger than the $4400 \AA$ flux
density, $R=S_{5 \mathrm{GHz}}/S_{4400 \AA}<10$ \citep{kel69}. So
any FIRST flux density detection or upper limit that implied $R<10$
was considered radio quiet with 1.4 GHz flux density used instead of
5 GHz flux density. We also eliminated the low ionization broad
absorption line quasars from our sample since they are known to have
anomalously weak [OIII]$\lambda$5007 and [OII]$\lambda$3727
emission lines \citep{bor92,zha10}. In the end, there were 6904
radio quiet sources remaining in our sample.

\par The spectral data of H$\beta$ and Mg II regime were separately reduced
using the procedures of \citet{Dong08, Wang09} which see for further
details, and we will only briefly outline it here. To measure the
H$\beta$ line and the [OIII]$\lambda$5007 line, a single power-law fit to the optical
continuum in the restframe wavelength range $4200-5600\AA$ was obtained taking into account
contributions from both broad and narrow Fe II multiplets that were modeled using the I Zw 1 templates
provided by \citet{veron04}. The H$\beta$ emission lines are modeled as multiple Gaussians: at
most four broad Gaussians and one narrow Gaussian with FWHM $<$ 900
km $s^{-1}$ for H$\beta$, and one or two Gaussians for each
[OIII]$\lambda$4959,5007. To measure the Mg II line, we fit the
continuum from the several continuum windows in the restframe
wavelength range 2200-3500\AA, after subtracting the estimated UV FeII contributions by
the Fe II template generated by \citet{tsu06} and the Balmer
continuum using the method by \citet{die02}. Each of the two Mg II
doublet lines is modeled with one broad five-parameter Gauss-Hermite
series component and one single Gaussian narrow component.
Furthermore, the broad components of the doublet lines are set to
have the same profile. The narrow components are set to FWHM $<$
900 km $s^{-1}$ and flux $<$ 10\% of the total Mg II flux. Finally,
we fit the spectrum in the [OII] regime using one Gaussian for [OII]
emission and a single power-law for optical continuum in the
restframe wavelength range 3600-3800\AA. This process resulted in five
pieces of relevant information for every quasar:
\begin{itemize}
\item The optical/UV continuum luminosity from 5100 $\AA$ to 3000 $\AA$, $L_{\mathrm{cont}}$
\item The luminosity of the broad component of H$\beta$ $\lambda$4861, $L_{H\beta}$
\item The luminosity of the broad component of Mg II $\lambda$2798, $L_{Mg II}$
\item The luminosity of the narrow line [OIII] $\lambda$5007, $L_{OIII}$
\item The luminosity of the narrow line [OII] $\lambda$3727, $L_{OII}$
\end{itemize}
\begin{figure*}
\includegraphics[scale=0.33]{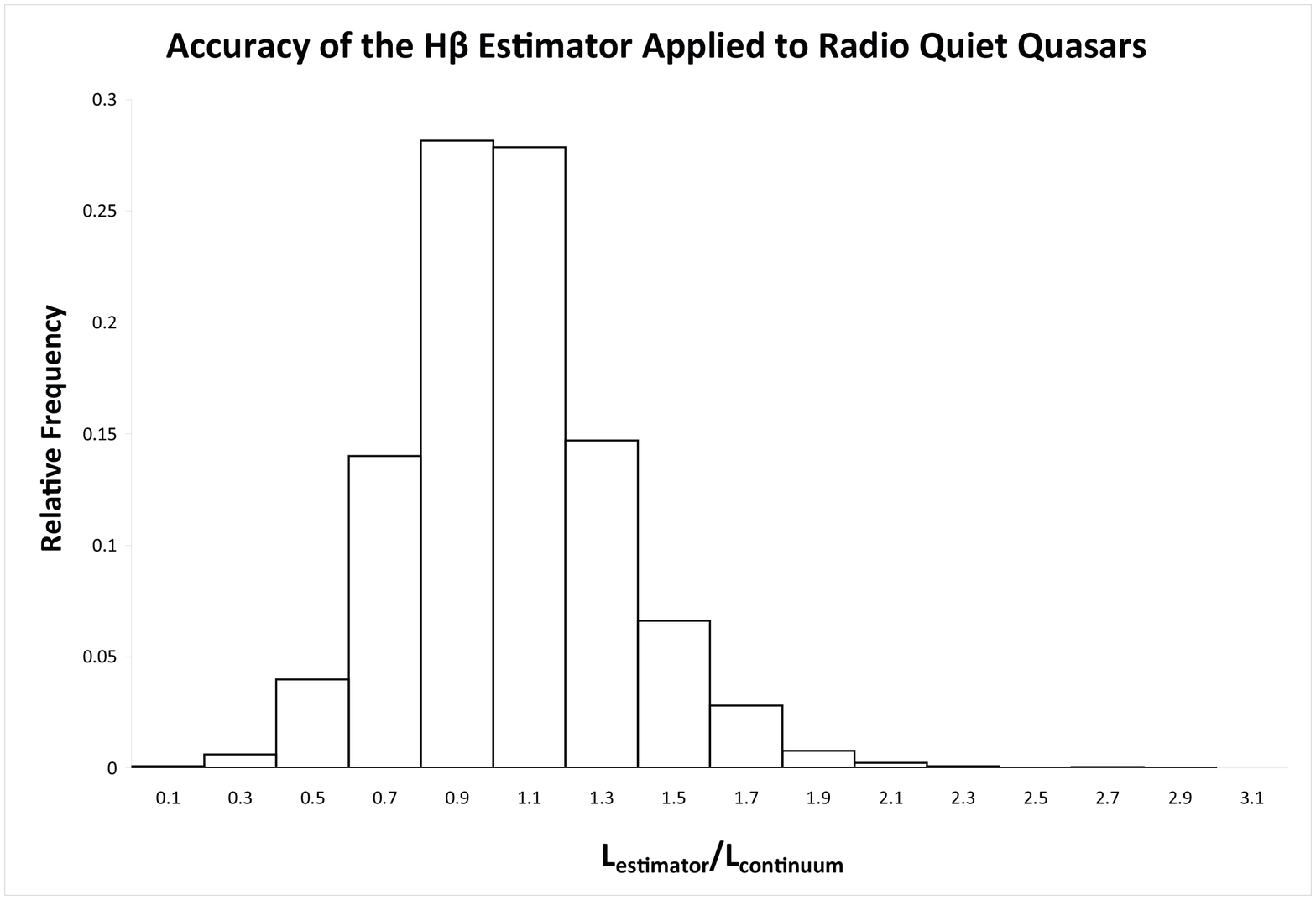}
\includegraphics[scale=0.33]{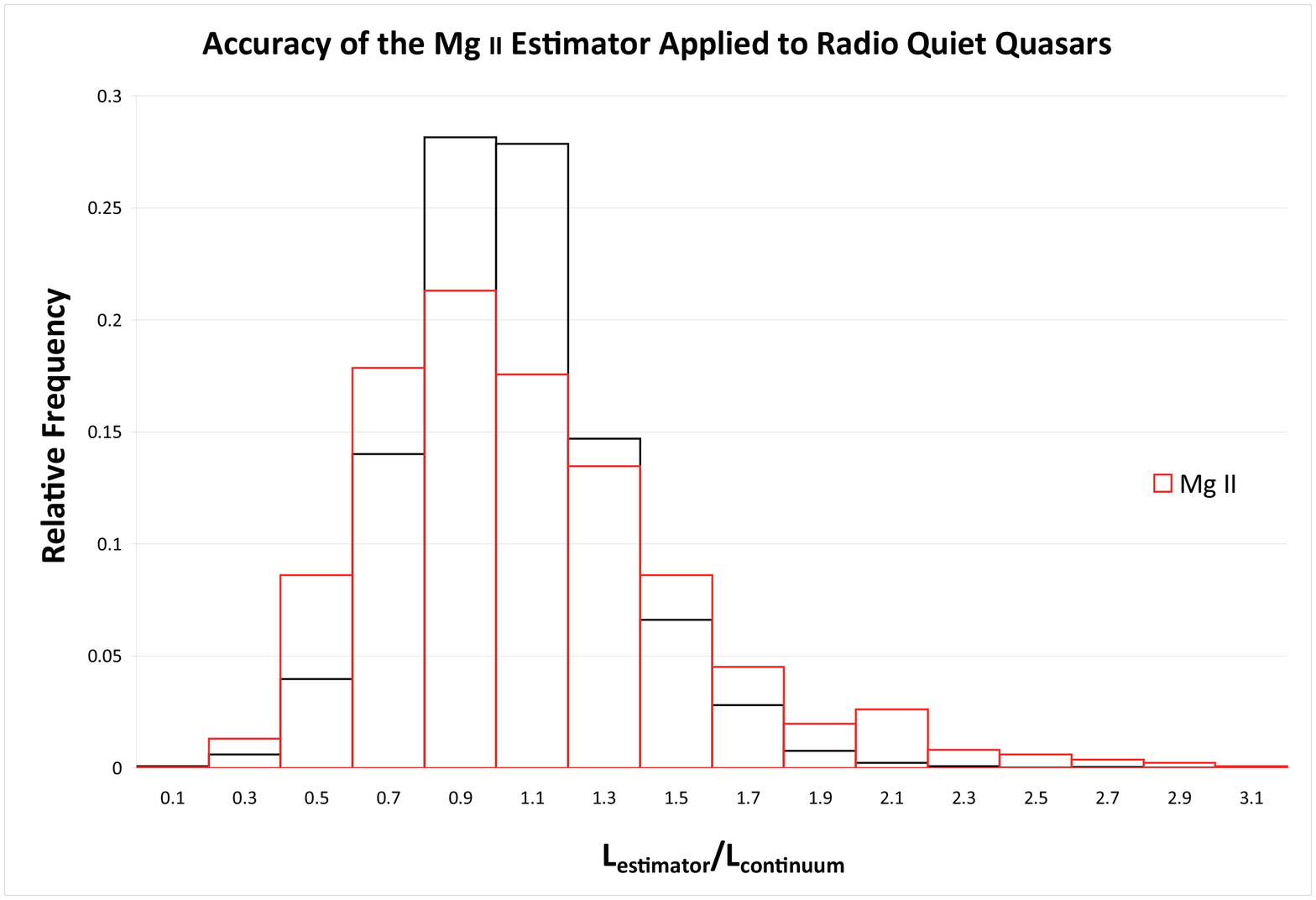}\\
\includegraphics[scale=0.33]{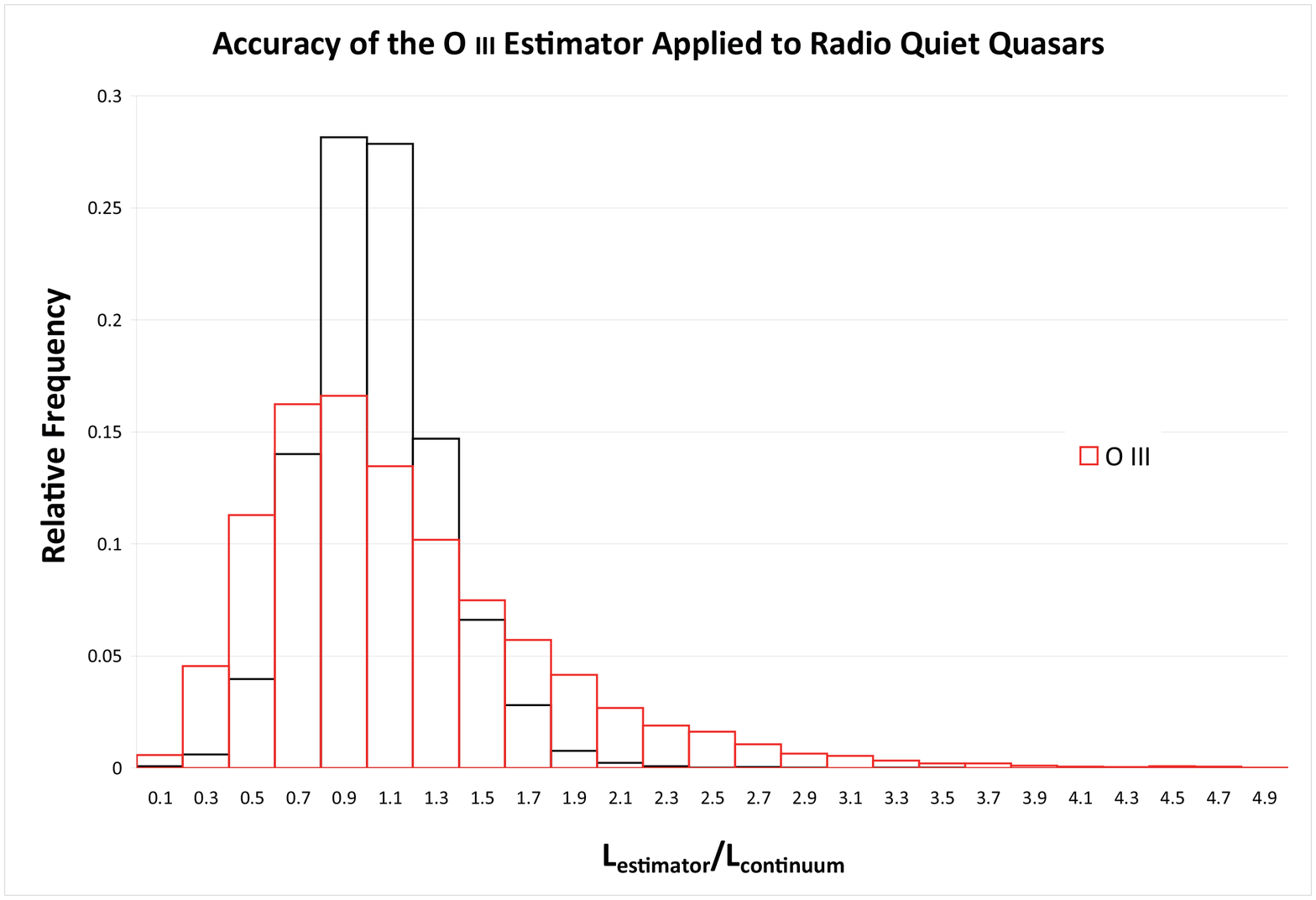}
\includegraphics[scale=0.33]{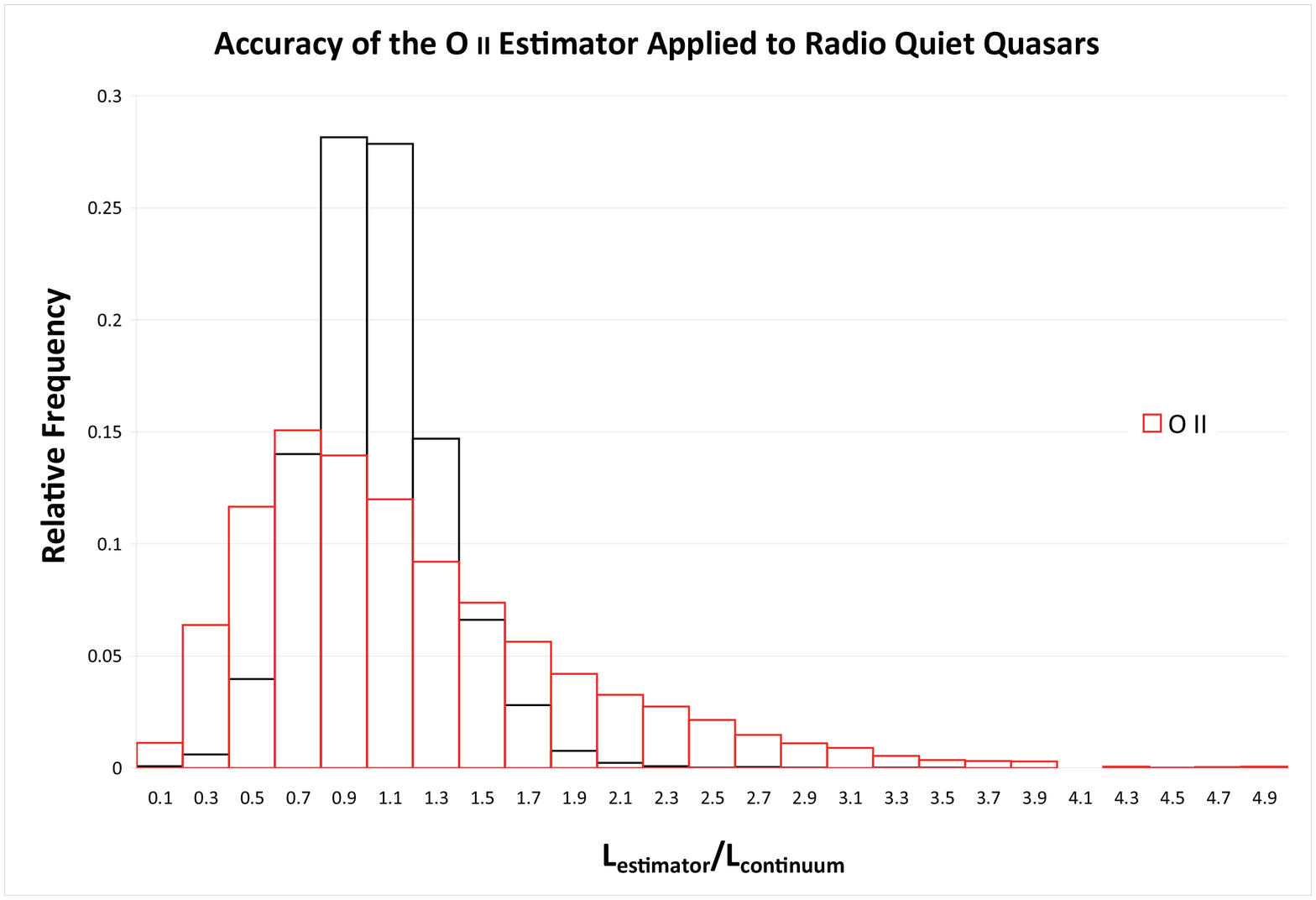}
\caption{The histograms
represent the distributions of the ratio of the estimated continuum
flux to the actual continuum flux,
$L_{\mathrm{estimated}}/L_{\mathrm{cont}}$ for each of the
emission lines in the context of the SDSS DR7 radio quiet subsample.
The smallest dispersion about 1 (the most accurate estimator) is
H$\beta$ in the upper left hand frame. The superior H$\beta$ fit is
shown in black in the other three frames for the sake of
comparison.}
\end{figure*}
\section{Line Strength Fits to Continuum Luminosity} The blue continuum luminosity is the most
basic signature of the thermal emission from the quasar, so it is
the most commonly used quantity for estimating $L_{bol}$. Perhaps
the most popular bolometric correction is the simple one proposed by \citet{kas00},
$L_{bol}\approx 9\lambda L_{\lambda}(5100\,\AA)$. Clearly, using a
portion of the optical/UV continuum is more accurate than a single
point and we have that at our disposal, $L_{\mathrm{cont}}$. For example, a single
point could lie in a noisy end of the SDSS spectrum. The
average spectral index in our sample of 6904 quasars, from 5100 $\AA$ to 3000 $\AA$, is
$\alpha_{\nu}=0.54$, where $\alpha_{\nu}$ is defined in terms of the
spectral luminosity as $L_{\nu} \sim \nu^{-\alpha}$. This spectral
slope implies that the \citet{kas00} bolometric correction can be
expressed as
\begin{equation}
L_{bol} \approx 15 L_{\mathrm{cont}}\;.
\end{equation}
Equation (1) is relatively insensitive to the choice of $\alpha$, with only a few percent change
in the constant of proportionality as $\alpha$ varies from 0.45 to 0.7.
\par Figure 1 shows how well the line strengths represent the continuum luminosity with the factor
of 15 bolometric correction from equation (1). The linear fits in
log-log space are given by equations (2) - (5) with the standard
errors for the intercept and the slope
\begin{eqnarray}
&& \log(L_{bol})= 12.32\pm 0.20 + (0.78 \pm 0.01)\log[L_{H\beta}]\;,\\
&& \log(L_{bol})= 16.76\pm 0.26 + (0.68\pm 0.01)\log[L_{MgII}]\;,\\
&& \log(L_{bol})= 26.50\pm 0.32 + (0.46\pm 0.01)\log[L_{OIII}]\;,\\
&& \log(L_{bol})= 33.96\pm 0.33 +(0.29\pm0.01)\log[L_{OII}]\;.
\end{eqnarray}
The coefficients of determination, $\mathbf{R}^{2}$, are shown on
the plots. The H$\beta$ fit is the best, followed closely by Mg II
with the narrow line fits considerably worse. Figure 1 confirms the
implications of \citet{sim98} that $L_{OIII}$ represents $L_{\mathrm{cont}}$
more reliably than $L_{OII}$ does. We also plot
archival estimators of $L_{bol}$ from the literature to show the
improvement obtained by these more rigorous calibrations of the
data. It appears that the \citet{kas00} normalization is
considerably higher than what has been used for line strength based estimates in
the past. A 5100 $\AA$ estimator based on the composite of
\citet{pun06}, $L_{bol}\approx 7.4\lambda L_{\lambda}(5100\,\AA)$,
is closer, but still considerably larger than the normalization of the archival line strength based estimates. The
\citet{mil92} estimator (which is about 65\% of the \citet{kas00}
value) seems to match the normalization of the line strength based estimates the
best. However, the poor archival fits are not just a consequence of
the normalization, but the slope of the fits as well. If the
normalization were adjusted, the \citet{wan04} fit for H$\beta$
would be fairly accurate.
\par There is no
unique criteria for the quality of the estimator and this must be
determined by each researcher per their requirements. We choose a
"factor of two" as a figure of merit (i.e. $0.5<
L_{\mathrm{estimated}}/L_{\mathrm{cont}}< 2$, where $L_{\mathrm{estimated}}$ is the
estimated continuum luminosity) for illustrative
purposes. The H$\beta$ based estimate is excellent with 97.4\% of the
estimates accurate to within a factor of two. The other estimators
are accurate to within a factor of two in the following order, Mg
II, [O III], [O II]; 90.1\%, 80.3\% and 73.9\% of the time,
respectively. The results are presented graphically in Figure 2
as the probability distribution (histogram) of the ratio of
estimated luminosity to measured continuum luminosity,
$L_{\mathrm{estimated}}/L_{\mathrm{cont}}$.
\begin{figure*}
\includegraphics[scale=0.33]{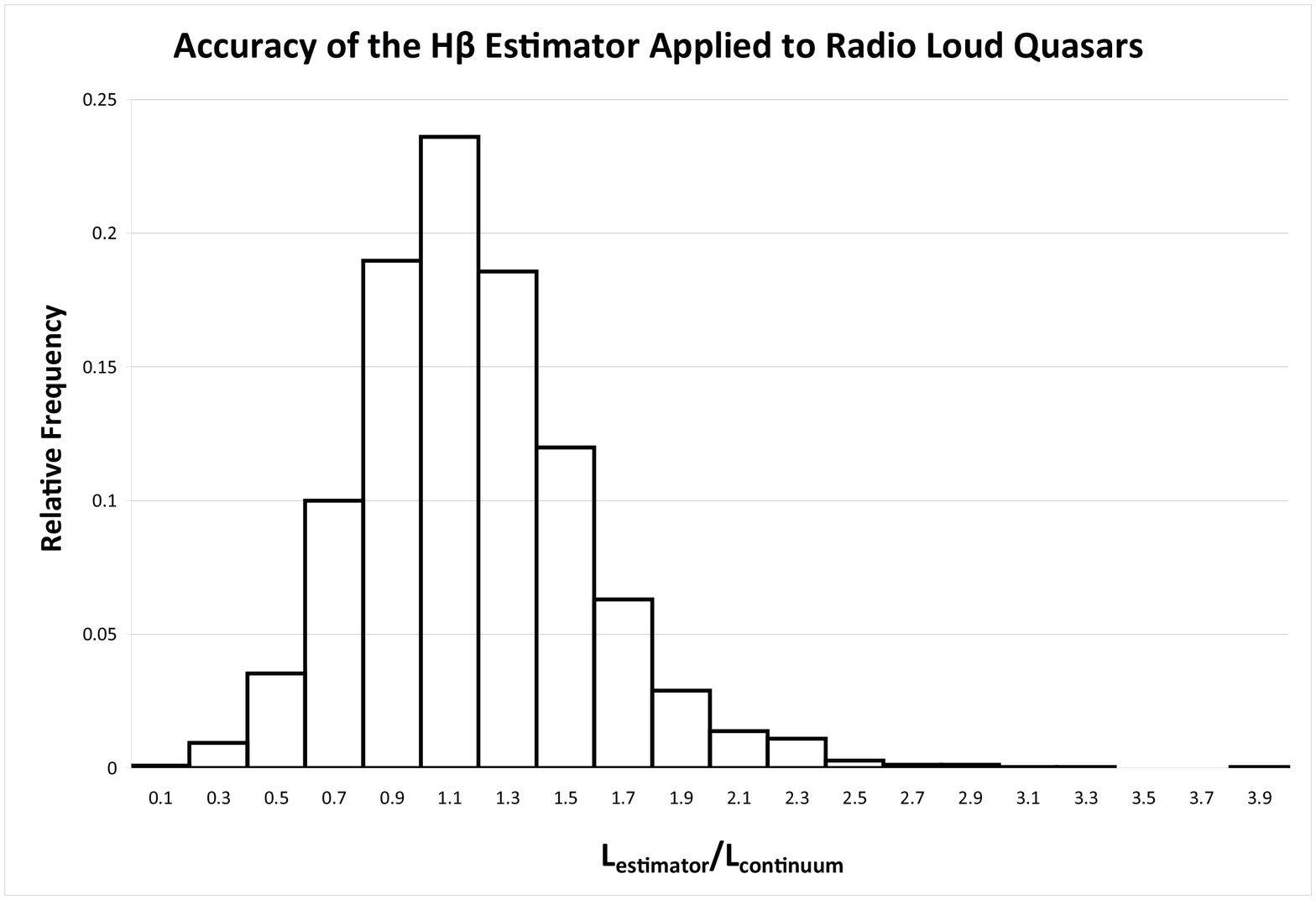}
\includegraphics[scale=0.33]{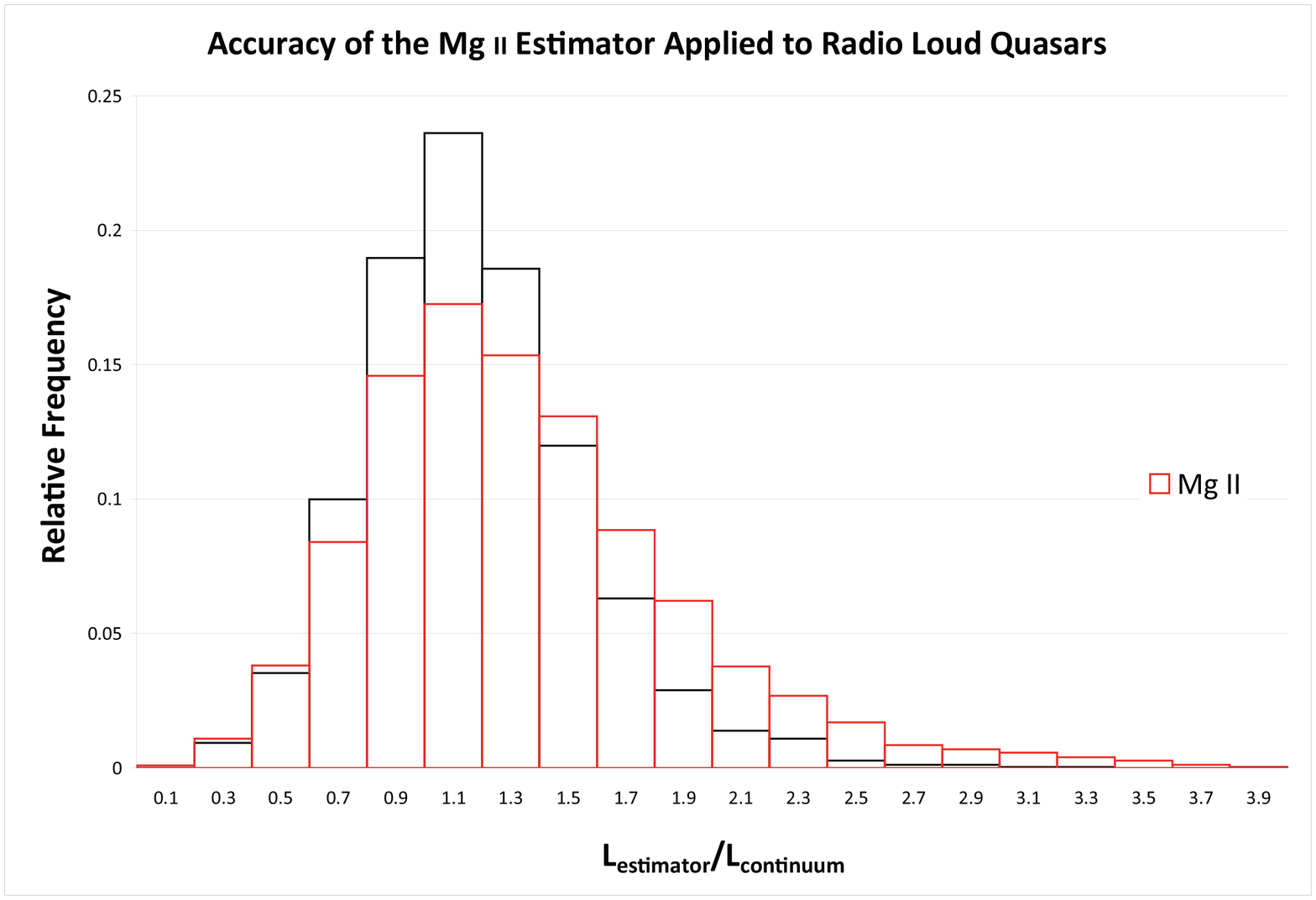}\\
\includegraphics[scale=0.33]{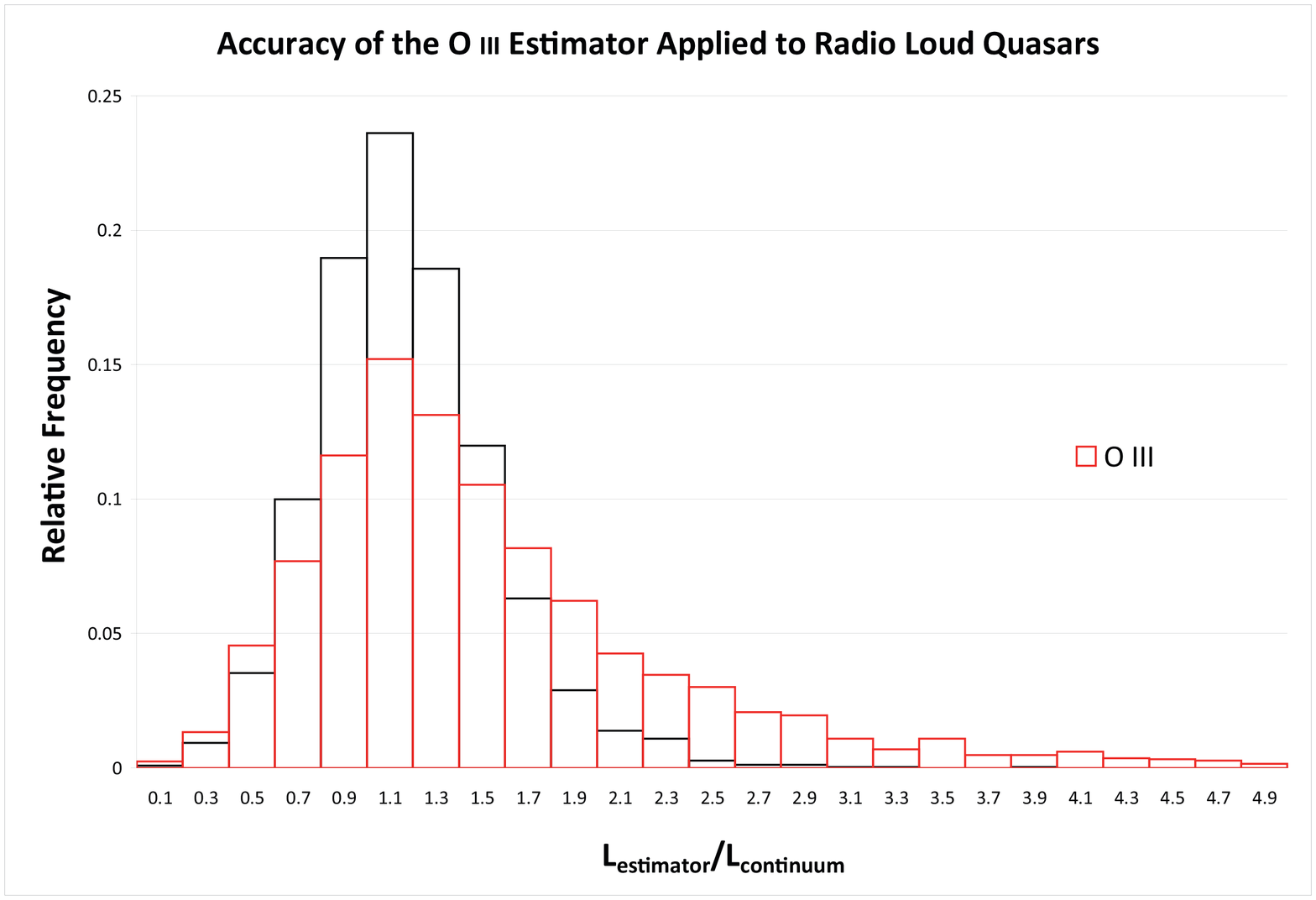}
\includegraphics[scale=0.33]{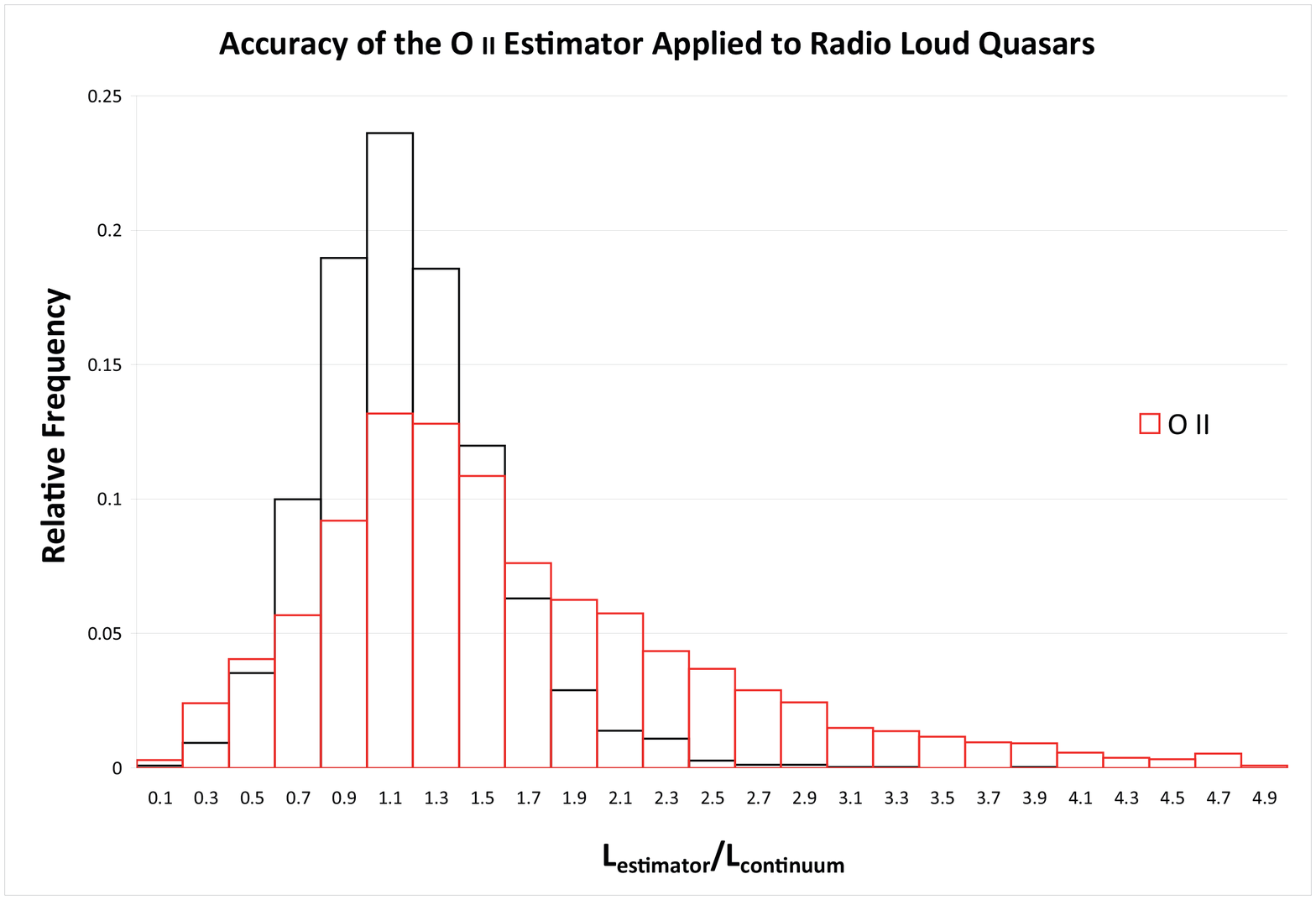}

\caption{The histograms
represent the distributions of the ratio of the estimated continuum
flux to the actual continuum flux,
$L_{\mathrm{estimated}}/L_{\mathrm{cont}}$ for each of the
emission lines in the context of the SDSS DR7 radio loud subsample.
The smallest dispersion about 1 (the most accurate estimator) is
produced by the H$\beta$ based estimate in the upper left hand frame. The superior H$\beta$ fit is
shown in black in the other three frames for the sake of
comparison.}
\end{figure*}

\section{Application to Radio Loud Quasars} There are 2461 quasars
in the SDSS DR7 sample, $0.4<z<0.8$, with a FIRST detection and
radio loudness, $R>10$. We use this sample as a test-bed for the
application of the estimators that were derived in the last section.
We can check our calibrations of the bolometric correction by
applying Equations (2) to (5) to each line individually and
comparing the resultant $L_{\mathrm{estimated}}$ to
$L_{\mathrm{cont}}$. The results are presented in Figure 3 as the
probability distribution (histogram) of the ratio of estimated
luminosity to measured continuum luminosity,
$L_{\mathrm{estimated}}/L_{\mathrm{cont}}$. Our "factor of two"
figure of merit (i.e. $0.5<
L_{\mathrm{estimated}}/L_{\mathrm{cont}}< 2$) indicates that the
H$\beta$ based estimate is still excellent with 94.9\% of the estimates
accurate to within a factor of two. The other estimators are
accurate to within a factor of two in the following order, Mg II, [O
III], [O II]; 86.2\%, 75.1\% and 67.5\% of the time, respectively.
Figure 3 shows that estimators based on the [OII] and [OIII]
emission line strengths have a propensity to over estimate the
continuum luminosity in radio loud quasars. This is in contrast to
the radio quiet quasars in Figure 2, where the more errant estimates
are equally likely to occur above or below "1". In other words, [O
III] and [O II] are often much stronger than expected from the
continuum luminosity alone. The implication is that the excess
narrow line strength is produced by the aforementioned excited gas
that is often observed aligned with the radio jets and is therefore
likely to be energized by the jet \citep{ink02,bes00,mcc95}. We
also performed the same exercise with a cutoff at $R>20$. The results
were very similar to those above, indicating that the definition
of radio loudness is a negligible factor in this analysis.
\par We give an example of how the line strengths can be used to
analyze radio loud AGN, the blazar 3C 216. We retrieve the line
strengths from \citet{law96}. For H$\beta$, Mg II, [O III] and [O
II], from equations (2) - (5), these line strengths yield the
following estimates for $L_{\mathrm{bol}}$, $8.12 \times 10^{44}$
ergs/sec, $1.23 \times 10^{45}$ ergs/sec, $7.24 \times 10^{45}$
ergs/sec, $1.32 \times 10^{46}$ ergs/sec, respectively. 3C 216 is
argued to be the one of the most kinetically dominated known
quasars, with a jet power (kinetic luminosity) to $L_{bol}$ ratio $>
10$ \citep{pun07}. The broad line estimates are in good agreement,
but the narrow line estimates are much larger. The results of this
paper indicate that the broad lines are therefore excellent
estimators of the continuum luminosity and the narrow lines are
dominated by a jet contribution.
\section{Conclusion}
We have derived line strength based estimators for $L_{bol}$ for
radio quiet quasars from the broad components of H$\beta$, Mg II,
and the narrow lines, [O III] and [O II] in Equations (2) -(5). The
strength of the broad component of H$\beta$ is a superior estimator
of $L_{\mathrm{cont}}$ for either radio quiet or radio loud quasars
since it is accurate to within a factor of two, 97.4\% and 94.9\% of
time, respectively. The next best line strength based estimators are
in order of accuracy, Mg II, [O III] and [O II]. We applied our
results to radio loud quasars and found strong evidence that the
narrow line based estimates are often skewed by what is likely a
strong jet induced contribution. It was also demonstrated that using
all four line strengths in tandem can be a useful diagnostic tool
for studying the jet power, environment and accretion power in AGN.
\par The results of this Letter are applicable to blazars, but perhaps not to NLRGs and
Seyfert 2 galaxies. There is compelling evidence that sometimes the
same gas that can attenuate the nucleus in these sources can also
attenuate the narrow line emission \citep{mul94,kra10}. In general,
without additional information, one does not know if the narrow line
region in a particular Seyfert 2 galaxy or NLRG is attenuated or
not, thus the [O III] and [O II] based estimators are not reliable in
isolation. We showed in section 4 that the narrow line estimators
are actually useful in the blazar context, where attenuation is not
an issue. In general, narrow line based estimators are most useful
if complemented by other information such as broad line luminosity
or the actual continuum luminosity. For NLRGs and Seyfert 2 galaxies, the IR continuum luminosity
is a valuable complement to the narrow line strengths since it is not believed to be attenuated \citep{fer10,ogl06}.

\end{document}